\documentclass{ptephy_v1}

\preprintnumber{XXXX-XXXX} 

\begin{document}

\title{Personal Recollection, 1945 - 1960}

\author{Toichiro Kinoshita}
\affil{Laboratory for Elementary-Particle Physics, Cornell University,
Ithaca, New York 14853, U. S. A. }
\affil{Amherst Center for Fundamental Interactions, University of Massachusetts, Amherst, Massachusetts 01003, U. S. A.}

\begin{abstract}%
This article gives a sketch of teachers and colleagues
who have had strong influence on my becoming a particle theorist.

\end{abstract}
\subjectindex{xxxx, xxx}

\maketitle

\section{Introduction}

The war ended on August 15, 1945.
Although large areas of Tokyo had been reduced to ashes by the intense bombing, Tokyo Imperial University (Todai) was luckily spared from destruction.
However, life in the Tokyo area had become very difficult, because of the extreme
shortage of housing and the breakdown of the food-rationing system,
established during the war, in the postwar confusion.
Todai was closed until the end of 1945, and students went back to
their homes in the countryside.

Todai reopened in January 1946.
I returned to Todai to resume my study of physics.
Tokyo was still extremely short of housing.
Some students and staff were living in their offices or laboratories.
Fortunately a cousin of my father offered me a room to live.

Sometime in the spring of 1946 Yoichiro Nambu showed up in the Physics Department
of Todai, still wearing his army uniform.
He had graduated from Todai in September 1942.
Then he served in the army until the end of the war.
But he was not discharged from the army for few more months,
since he was in charge of dismantling the army research facility to which
he was assigned.

I had enrolled in the Physics Department of Todai in September 1944.
Because of the war-time measure of accelerated curriculum our class
took all of our first-year courses during the fall semester of 
1944 and all of our second-year courses
during the spring semester of 1945.
Thus when I returned to Todai in January 1946, there were no more
formal courses to take.
I was able to work for more than one and a half
years on research projects without being bothered by course
work or exams.

Before describing the early post-war period,
I should like to mention an important work of S. Tomonaga
published in 1943 \cite{Tomonaga43}.

During the war most physicists were mobilized to the war effort.
Tomonaga was assigned to develop radar.
I understand that
he developed an elegant theory of radar based on the idea of the S-matrix,
to which he was exposed while visiting Heisenberg before the war.
What is more relevant for the development of physics in post-war Japan,
Tomonaga found time to work on the super-many-time reformulation
of quantum field theory \cite{Tomonaga43}.
This was an attempt to reformulate the non-relativistic form of the
Heisenberg-Pauli theory \cite{H-P29} into a fully relativistic form.

\section{ Kodaira seminar}

I was interested in particle physics, or quantum field theory in
general.
But in 1946 Todai had no professor of particle physics.
Looking for an advisor, I recalled that I was strongly impressed  
by the extreme clarity of the lectures on mathematical physics 
Professor Kunihiko Kodaira gave during the fall term of 1944.
Although he was not a particle physicist, I asked him to supervise my
study of quantum field theory.
He kindly agreed to supervise me and my classmate Yoichi Fujimoto in the study of particle physics.

Actually, although Todai had no professor of particle physics at that time,
there was no shortage of people who were interested 
in particle physics. 
Discussion of particle physics had been going on all the time.
Particularly active were Ziro Koba, Satio Hayakawa,
Yoichiro Nambu, Gyo Takeda, Yoshio Yamaguchi, and Yoichi Fujimoto.

When Prof. Kodaira asked us what we wanted to study,
we said that we wanted to begin with the latest papers available
on quantum field theory.
(This is because we had studied by ourselves many of the 
early papers on quantum field theory.)

During the war research papers from abroad stopped coming to Japan.
But we did find some fragments of journals published
 during 1941 - 1944 in the library of the Todai physics department.
My memory is not clear but I think we chose as the first topic of our
weekly seminar Dirac's classical theory of the electron \cite{Dirac42} 
or his theory of the indefinite metric \cite{Dirac43}.

In a footnote of one of these papers, we found that Pauli mentioned
that ``the S-matrix theory of Heisenberg is a picture frame without a
picture" \cite{PtoD43}.
From this footnote we learned that Heisenberg wrote two papers
on the S-matrix, and we began looking for them.

According to my classmate Y. Yamaguchi
Zeitschrift f\"{u}r Physik journals containing these articles were brought to
Japan during the war by a German submarine, and some copies were
kept in the office of Professor Tomonaga at Tokyo Education University.
Yamaguchi went to Prof. Tomonaga, and brought copies back
to Todai with  his permission.
In view of the ``Top Secret" stamp on the covers of these copies,
their existence must have been known to only a few people
around Prof. Tomonaga.
We chose these papers \cite{Heis43a} and 
\cite{Heis43b} as the subject of our next
seminar and studied them intensely.

Quantum electrodynamics (QED) was formulated around 1929 by Heisenberg,
Pauli, Dirac, and others.
However, for many years QED suffered from a curious disease:
Although it agreed well with experiments in the lowest
order of perturbation theory, attempts to improve agreement further
by calculating higher-order corrections always ran into
divergent results.
Clearly something was wrong with QED.

Since the 1930's many people had tried to solve this puzzle.
Heisenberg's S-matrix papers may be regarded as one of these efforts.
His first paper is an attempt to realize his philosophy that the divergence
difficulty of quantum field theory must be solved by describing
physical phenomena in terms of observable quantities only.

Specifically, he was trying to remove the Hamiltonian from the theory,
because he regarded it as unobservable.
The first paper \cite{Heis43a} presents only the framework of a theory,
as Pauli had observed.
The second paper \cite{Heis43b} is an attempt to put a ``toy picture" in it.
Heisenberg was trying to understand the mechanism of cosmic-ray
showers, to see whether they are multiple production
(many particles are produced in one collision) or
plural production (particles are produced in several successive
 collisions),
which was then a hot topic
among cosmic-ray physicists.

This model is not a successful theory as far as physics is concerned.
However, Heisenberg explored various consequences of this model
in great detail.
His calculation was rather difficult for our shallow knowledge
of mathematics.
Prof. Kodaira showed us how to treat these problems.
At some point it was necessary for us to study the behavior of
a function with two complex variables.
We were ignorant of the theory of functions of several complex variables.
We were fascinated by Prof. Kodaira's description of the Riemann surface
of 2 complex variables as a picture of 4-dimensional space of
real variables.

Prof. Kodaira became interested in Heisenberg's S-Matrix theory
not as a theory of physics but because of its mathematical
structure.
He realized that Heisenberg's method is actually related to
the eigenvalue problem of the Jacobi matrix.
Thus we began to study the Jacobi matrix.
Since the last chapter of M. H. Stone's book ``Linear Transformation in Hilbert Space" 
has an introduction to the Jacobi matrix,
we students studied this chapter.
But to understand this chapter it was necessary to understand
the preceding chapter, and then further earlier chapters.
By the time we arrived at the chapter on 2nd-order ordinary 
differential equations,
Prof. Kodaira had discovered a general formula for eigenfunction
expansion of 2nd-order ordinary differential equations,
and we students were prepared to understand his  new theory.

Tokyo in 1946 was still burned out and short of food.
Circumstances were far from ideal for studying physics.
However, we worked hard to understand the Heisenberg's theory
step by step.
The seminar started right after lunch.
Occasionally it lasted past 8.00 pm without a supper break.
For a short rest at around 3.00 pm Prof. Kodaira took out
a lemon, partly covered by mold,
 from the drawer of his desk 
and put a small slice in a potful of black tea.
I still remember its taste.
 
Prof. Kodaira summarized his work in a preprint,
``The eigenvalue problem for ordinary differential equation
of the second order and Heisenberg's theory of S-matrix",
and asked Prof. Yukawa, who happened to be going to 
the Institute for Advanced Study in Princeton,
to give it to H. Weyl.
Shortly afterwards, Kodaira received a letter from Weyl
informing him that Titchmarsh had obtained the same result
by an entirely different method.
Furthermore, Weyl kindly sent Titchmarsh's book to Kodaira.
We found that Titchmarsh's method was somewhat pedestrian,
in contrast to Kodaira's very 
elegant method which
used Weyl's ``limit circle" method effectively and 
displayed a deep understanding of the Hilbert space.

My memory is hazy whether it was in late 1946 or early 1947,
but it became widely known  around Tokyo
 that Heisenberg's S-matrix theory
was being studied by the Kodaira group,
and Fujimoto and I were asked to give an open seminar on 
Heisenberg's S-matrix theory.
We were surprised to find that the lecture room was 
overflowing with people.
We managed to present the seminar without embarrassing mistakes.
According to S. Hayakawa (2 years senior to our class),
Professor Tomonaga attended this seminar.
But I do not recall seeing him in the audience.
Probably I was too nervous in our first public lecture.

\section{ Tomonaga seminar}

Shortly after the end of the war
Prof. Tomonaga assembled several 
physicists from Tokyo Education University
(predecessor of today's Tsukuba University) and Todai,
and began to extend his super-many-time theory to QED and other
field theories.
His seminar was held in a building standing in the middle of
the burned-out ruins of an army facility at Okubo in northern Tokyo,
because his own University had been destroyed by bombing.

Sometime toward the end of 1946 or early 1947 Y. Yamaguchi
took Y. Fujimoto and me to visit Tomonaga's
lab in Okubo, which was about one hour's trip away from Todai.
Tomonaga agreed to let us join his weekly seminar.

Tomonaga's seminar became very popular
and attended by more than 30 people from the Tokyo area.
The main texts used in his seminar were Heitler's
``Quantum Theory of Radiation" and Bethe's review article
on Nuclear Physics published in
the Reviews of Modern Physics.
Besides discussing these texts, we often went back to
the original articles to clarify questions.
I recall in particular the article of Bloch and Nordsieck \cite{B-N37}
concerning the infrared (IR) catastrophe, and
the article of Pauli and Fierz \cite{P-F39} analyzing the ultraviolet
(UV) divergence.

In many countries research in physics, interrupted by the war,
was resumed and new discoveries began to be published.
New literature was not yet available in the library
of Todai's Physics Department.
However, the latest issues of the Physical Review, as well as
magazines such as NEWSWEEK and TIME, began to arrive at the CIE Library
(of the Occupation Force) at Hibiya, not far from the Tokyo Central Railroad Station.
We used to visit this library early in the morning
so that we did not miss the latest journals and magazines.
This is where we first learned about the 
Shelter Island Conference held in the summer of 1947
where the discoveries of the Lamb shift
\cite{L-R47} and the anomalous magnetic moment of the electron \cite{K-F47} were reported.

When the exciting news from the U. S. arrived, the Tomonaga group
was ready to tackle these new developments.
Within few weeks Tomonaga and Koba discovered the theory of
renormalization \cite{T-K47}, which removed the major difficulty of UV divergence
of quantum electrodynamics (QED) and opened a way for new
developments of quantum field theories.

I recall that I had difficulty understanding
Weisskopf's calculation of the ``electron's self-energy" 
\cite{Weisskopf39}
quoted in Heitler's text,
and asked Prof. Tomonaga about it.
Instead of answering my question, he told me to give two or three seminars on Weisskopf's paper.
Weisskopf calculated the second-order radiative correction
to the self-energy of the electron by exactly following the method of
Dirac's hole theory. 
Verifying his calculation step by step with great care,
I found that my questions were all based on my misunderstanding
and Weisskopf's result was in fact correct.
This experience was very helpful when I read Dyson's article
later \cite{Dyson49}.
The self-energy of an electron is infinitely large
and was regarded as one of the difficulties of QED.
What Weisskopf found is that, the divergence is not as
strong as linear or quadratic but is actually a very mild
logarithmic divergence.
If the divergence were stronger than logarithmic,
the renormalization method would not have succeeded. 
In other words Weisskopf's result was
a crucial step towards the success of renormalization theory.

In the U. S., Schwinger \cite{Schwinger48a} discovered the renormalization of QED and
applied it in particular to the calculation of the anomalous magnetic moment
of the electron \cite{Schwinger48b}.

Nambu worked on the electron anomalous magnetic moment, too.
Unfortunately, he did not publish his work since Schwinger
published his paper first.

Bethe \cite{Bethe47} solved the Lamb shift problem,
applying the renormalization idea to the non-relativistic version of QED.

In August 1948, Oppenheimer invited Yukawa to the Institute
for Advanced Study in Princeton.
Yukawa used this opportunity to transmit by numerous letters 
the exciting new developments
in the U. S. to physicists in Japan.
In particular, we were inspired by the activity of young
physicists such as Feynman and Dyson.

\section{ Collaboration with Koba}

At the beginning of 1948 there were two competing methods to deal with
the ultraviolet divergence of the electron self-energy: one is Tomonaga's renormalization method
and the other is the C-meson theory proposed by Prof. Sakata
of Nagoya University \cite{Sakata47},
which cancels the UV divergence due to the radiative correction
provided that
\begin{equation}\label{eq1}
f^2 = 2 e^2,
\end{equation}
where $f$ is the coupling constant of the C-meson to the electron.

Naturally the question was raised as how to decide experimentally
which one was correct.
I asked Prof. Tomonaga for his opinion.
He told me that Z. Koba was working on this problem right now,
so that it would be a good idea to work with him. 
In collaboration with Koba, I began to study the difference between
the predictions of C-meson theory and renormalization theory. 
What we found may be summarized as follows: 

\begin{itemize}

\item  The predictions of Tomonaga's theory and the {\underline {original}} form of
Sakata's theory can be 
distinguished by examining the elastic scattering of an electron
by a fixed potential for an incident electron energy of up to several
hundred Mev \cite{E-K-K48}.

\item Shortly after publication of that result, however, it occurred to us that the renormalization must be applied to the C-meson theory, too.
 When renormalization is applied to the C-meson theory,
 the two theories become
    indistinguishable within the experimental precision
    of the elastic scattering up to 
    several hundred Mev (highest energy then available) \cite{E-K-K49}.
\end{itemize}

Several years later I proved that C-meson theory
does not remove the divergence of self energy in the 4th order.
This was the end of C-meson theory \cite{Kinoshita50a}.

After this collaboration Koba moved to the Yukawa Institute of Theoretical
Physics at Kyoto University.
Thus I did not have any further occasion to work with him.
He later moved to Poland, and then to the Bohr Institute in Copenhagen.
Unfortunately, he passed away prematurely in 1973.

\section{ Infrared divergences in the Feynman-Dyson theory}

Bloch and Nordsieck treated the infrared problem by a non-perturbative
method assuming that all infrared photons have zero energy \cite{B-N37}.
However, their method cannot be applied directly to the emission and
absorption of real photons of non-zero energies.
In order to deal with this problem 
I developed a method to handle the infrared 
problem within the perturbative framework of Feynman-Dyson theory \cite{Kinoshita50b}.
My insight was not deep enough, however, and I overlooked a subtlety in the infrared problem.
This became the source of an error in the calculation of radiative correction to the $\mu$-e decay,
as is described in detail in Sec. 7.

\section{ Collaboration with Nambu}

In 1948 I was trying to see whether the renormalization
theory of Tomonaga could be extended to the interaction of a charged
vector meson with the electromagnetic field.
I found out that Nambu was also interested in the same problem.
Thus we started working together.
While our work was in progress, we received from Prof. Yukawa
a brand new preprint of Dyson's paper \cite{Dyson49},
which proved the equivalence of
the Tomonaga-Schwinger theory and Feynman's theory,
in spite of their very different appearance.
(Actually at that time we did not know what Feynman's theory looked like,
since Feynman did not publish his theory until a year later \cite{Feynman49}.
Thus we had to guess what Feynman's theory looked like
by studying Dyson's paper.)
The Feynman-Dyson theory opened up a new method of computation,
that was much more transparent and systematic than the perturbation
method based on the non-covariant formulation 
of Heisenberg and Pauli \cite{H-P29}.

Recognizing the power of Feynman-Dyson theory, Nambu and I quickly
reorganized our theory a l\`{a} Dyson.
Also, instead of the conventional formalism of vector mesons,
we adopted the Duffin-Kemmer equation \cite{Kemmer39} as the starting point to
take advantage of its close structural similarity with the Dirac equation.

In April of 1949 Nambu, together with S. Hayakawa, Y. Yamaguchi,
and K.Nishijima, moved to the newly established Physics
Department of Osaka City University.
Thus my collaboration with Nambu came to an end.
In spite of this exodus the particle physics group at Todai remained very active with
younger members such as H. Miyazawa and M. Koshiba.

In the summer of 1949 both Tomonaga and Kodaira were invited to
the Institute for Advanced Study in Princeton.
Tomonaga returned to Japan next year so that the Tomonaga seminar
continued uninterrupted during his absence.
The Kodaira seminar was terminated, however, since he stayed
in the U. S. for many years.

Nambu and I wrote up papers on our work \cite{K-N50a, Kinoshita50, K-N50b}, and sent them to
were invited to the Institute for Advanced Study
for the academic year 1952 - 1953 (extended to 1954).
This opened up a possibility to collaborate with Nambu again.

I left Japan from Nagasaki in early August 1952 
by a cargo boat, and arrived at Seattle about 17 days later.
Then I moved to Berkeley by Greyhound bus, and visited Berkeley,
Caltech, and UCLA while waiting for Nambu's arrival.
Nambu had left from Yokohama by a cargo boat and arrived at San Francisco
about two weeks later.
We took a train named ``California Zephyr" to cross the Rockies,
stopping at Denver to visit the cosmic-ray laboratory at
Mt. Evans, elevation 14,000 ft.  The next stop was at Chicago
to visit the University of Chicago and talk with the people who found
the 3-3 resonance of pion-nucleon scattering
with their proton cyclotron. 
 Then we stopped at Rochester to visit R. Marshak at the University 
of Rochester, and finally arrived at Princeton in late August.

At the Institute for Advanced Study
 Nambu and I were given desks in the same room on the second
floor of the physics building.
This room was farthest from the office occupied by Lee and Yang.
But when they started arguing we could hear them very loudly
from our office.
We tried to develop a general theory of Hartree fields,
a collective description of many-particle systems \cite{K-N54}.

At the end of the Spring term of 1954 Nambu was invited to the University
of Chicago, where he used our paper \cite{K-N54}
as the starting point for his theory of superconductivity,
which was developed further to the theory of spontaneous symmetry
breaking.

In May of 1954 I became puzzled by the sizable discrepancy
between theory and measurement of the ionization energy of the He atom,
and began examining  Hylleraas' well-known calculation.
This work continued for several years at Columbia (1954 - 1955) and Cornell (1955 -).
At last I found that the discrepancy was caused by a subtle
error in Hylleraas's numerical work \cite{Kinoshita59a},
which did not strictly follow the rules of
the variational method.

I was told that Fermi was intrigued by this problem and conjectured
that the He atom might have some new force acting between
electron and nucleus.
To my disappointment my calculation disproved his conjecture.

I learned a lot from Nambu in our collaboration.
But we pursued different paths afterwards.

\section{ Radiative Correction to the $\mu - e$ decay}

The year 1956 was a very exciting year. 
After an exhaustive examination of the so-called $\theta$-$\tau$ puzzle,
T. D. Lee and C. N. Yang suggested that
the parity symmetry is broken in the weak interaction
and proposed ways to test it experimentally \cite{L-Y56}.
Experimental verification followed soon afterwards 
\cite{Wuetal57, Garwinetal57}.
The two-component neutrino theory became the favored theory.

A detailed comparison of the two-component theory with experiment
would require inclusion of radiative corrections.
Thus I started working on the radiative correction to 
the parity-violating part of the $\mu - e$ decay.
It turned out that Albert Sirlin had worked on the radiative correction to the parity-conserving muon decay before coming to Cornell
as a graduate student \cite{B-F-S56}.
In collaboration with him I calculated the radiative correction to the
parity-violating part of the muon decay.
As expected, the infrared divergent terms of real and virtual emissions
of photons canceled out so that
the radiative correction to the decay lifetime of muon was
finite, although it was very large: \cite{K-S57a}
\begin{equation}\label{eq2}
\frac{\tau-\tau_0}{\tau_0} = - \frac{\alpha}{2\pi}
 (\omega^2 + ...)  \simeq - 0.02973...
\end{equation}
where $\omega = \ln (m_\mu / m_e) = 5.3316 ...$.

We presented our result at the Rochester conference in the spring of
1957.
Then, one day in 1957 or 1958, we received a preprint from 
S. Berman stating that the paper 
\cite{K-S57a} was wrong because of an inconsistent treatment of
real and virtual emissions of photons.
(See next paragraph for more details.)
Correcting this error he obtained an answer that is linear in $\omega$.
When we read his preprint, however, we suspected immediately that
his result must also be wrong
since the mechanism to remove the $\omega^2$ term
is likely to remove the $\omega$ term, too.  Alberto and I worked hard for a week or two and found that
our intuition was in fact correct:
No term proportional to $\omega$ or 
$\omega^2$ is present in the muon life-time \cite{K-S59}.
Berman agreed with us and revised his paper accordingly \cite{Berman58a}. 

The error in \cite{K-S57a} was caused by the incorrect treatment of
real photons. 
Recall that, in a covariant calculation,
the virtual photon is treated as a vector meson of mass $\lambda$
with the understanding that $\lambda \rightarrow 0$
in the physical limit.
To be consistent with this, the real photon must also be regarded as a vector meson of mass $\lambda$.
This means in particular that the real photon has longitudinal
and time-like polarizations in addition to two transverse polarizations.
Our mistake was to consider only the contribution of
transverse polarizations. 
The contribution of longitudinal and time-like polarizations does not
vanish if the photon has infrared divergence, even in the limit
$\lambda \rightarrow 0$.
As a matter of fact this extra contribution has an $\omega$ dependence
that cancels exactly the $\omega$ dependence from transverse photons.

Feynman visited Cornell for three months in the fall semester of 1958.
He explained to me how he and Berman made exactly the same mistake.
Feynman had asked Berman to check our calculation for his thesis
work. Actually, Feynman told me that he 
himself was doing this calculation 
independently in order to check Berman's work.
At the end they compared notes and were satisfied that their results
agreed.
When confronted with our new result, they checked notes once again and found that they made the same mistake in copying the
bottom equation of a page to the top of next page.
Feynman was so disturbed by this mistake that he told me
how sorry he was more than a few times while at Cornell.

See \cite{Kinoshita03} for more detailed discussion of this section.

\section{ Radiative Corrections to $\pi - e$ and $\pi - \mu$ decays}

Feynman brought to Cornell a new preprint by Berman on the 
one-loop radiative correction to $\pi - e$ decay.
Berman's result was of the form
\cite{Berman58b}
\begin{equation}\label{eq3}
\frac{R}{R_0} = 1 - \frac{3\alpha}{\pi} \ln \left ( \frac{m_\mu}{m_e} \right ) + ...
\end{equation}
where
\begin{equation}\label{eq4}
R_0 = \left (\frac{m_e}{m_\mu} \right )^2 \left ( \frac{m_\pi^2 - m_e^2}{m_\pi^2 - m_\mu^2} \right )^2
\end{equation}
is the uncorrected ratio of $\pi-e$ and $\pi -\mu$ decay rates in the V-A theory.

The value of this $R$ looked strange since $R/R_0$ diverges for 
$m_e/m_\mu \rightarrow 0$.
Feynman and I were so puzzled by this result that we decided to
check it with a fresh calculation.

For the next two months we worked hard, totally independent of
each other, except that we agreed to start from the same effective Lagrangian
\begin{equation}\label{eq5}
g m_l \bar{\psi}_l a \psi_\nu \phi_\pi,
\end{equation}
where $l$ represents the physical mass of muon or electron and $a=(1+i\gamma_5)/2$.

The most time-consuming part of the calculation is integration over
the final states consisting of an electron, a neutrino, and a photon.
Following the conventional method it took for me
about two months to evaluate this integral.
The calculation occupied over 30 sheets of papers.
Although Feynman spent about the same amount of time,
he was looking for a more efficient method and found one.
His approach is to maintain Lorentz invariance throughout
the intermediate steps of calculation, 
which allowed him to choose a frame of reference 
in which angular integration becomes trivial.
To my great amazement
this enabled him to write 
the complete derivation in just two sheets of paper.

When we finished the calculation and 
compared the notes we found that
we agreed with each other;
in particular, our results did not contain the $\ln (m_\mu /m_e)$ term.
We thought for a while that Berman's calculation was wrong.
But, after scrutinizing our calculation very closely,
we realized that it was our effective Lagrangian (\ref{eq5}) that
was not justifiable.

Berman started from an effective Lagrangian containing a derivative
coupling:
\begin{equation}\label{eq6}
g \bar{\psi}_l\gamma_\mu  a \psi_\nu \left ( i \frac{\partial \phi_\pi}{\partial x_\mu}
-e A^\mu \phi_\pi\right ).
\end{equation}
This term is to be evaluated with respect to the states
which include radiative corrections up to the order $e^2$.
To compare (\ref{eq6}) with (\ref{eq5}) turn it into a non-derivative form
\begin{equation}\label{eq7}
g( m_l^0 - m_\nu^0) \bar{\psi}_l a \psi_\nu \phi_\pi,
\end{equation}
by integration by part, and using the equation of motion
\begin{equation}\label{eq8}
i \frac{\partial}{\partial x_\mu} \bar{\psi}_l \gamma_\mu
+ e A^\mu \bar{\psi}_l \gamma_\mu 
+ m_l \bar{\psi}_l
- \delta m_l \bar{\psi}_l = 0,
\end{equation}
where $m_l^0$ and $m_\nu^0$ are {\bf bare} masses,
$m_l$ is the physical mass,
and $- \delta m_l \bar{\psi}_l$ is the counterterm to the radiative correction.
The relation $m_l = m_l^0 + \delta m_l$ was used in Eq. (\ref{eq7}).

It is seen that when the radiative correction (to order $e^2$) is included
in the same way as in the ordinary QED, Eq. (\ref{eq7}) depends on
the bare mass $m_l^0$ instead of the physical mass.
The appearance of bare mass in a similar context had been noticed by
Ruderman \cite{R-W56}.
On the other hand, the
Lagrangian (\ref{eq5}) was obtained from the equivalent Lagrangian (\ref{eq7})
assuming that $m_\nu^0 = 0$ and $m_l^0$  is the physical lepton mass.
The discussion above shows that
it was this assumption that was not justifiable.

Eq. (\ref{eq7}) shows that
Berman's Lagrangian may be written as (assuming $m_\nu^0 = 0$)
\begin{equation}\label{eq9}
g m_l^0 \bar{\psi}_l a \psi_\nu \phi_\pi.
\end{equation}
Namely, our result can be turned into Berman's result
by simply replacing the physical mass in (\ref{eq5}) by the bare mass.

At first sight it would look strange that the $\ln (m_\mu/m_e)$ term
appears in Eq. (\ref{eq3}) in view of absence of such a term in 
the total probability
according to the mass singularity theorem \cite{Kinoshita62}.
Actually Eq. (\ref{eq3}) is consistent with the theorem
since it is multiplied by $m_e^2$ as is seen from Eq. (\ref{eq4})
and thus vanishes in the limit $m_e \rightarrow 0$.

Although Feynman was fully involved in this calculation,
he refused to put his name on the paper \cite{Kinoshita59b}.

The calculation in \cite{Berman58b}
and \cite{Kinoshita59b} relied on implicit untested assumption
that the effective Lagrangians (\ref{eq5}) and (\ref{eq6}) are viable
and that the UV cut-off is common for both $\pi-e$ and $\pi-\mu$ decays.
Justification of these assumptions had to wait for the emergence
of the standard model calculation \cite{M-S76},
which favors Berman's result.
Taking the hadronic effect into account \cite{M-S93},
the value of $R$ becomes
\begin{equation}\label{eq10}
R = 1.2352 (5) \times 10^{-4}.
\end{equation}
This is in good agreement with the experiments \cite{Brittonetal92,
Czapeketal93}:
\begin{equation}\label{eq11}
R = 1.2265 (34) (44) \times 10^{-4},~~~~~~
R = 1.2346 (35) (36) \times 10^{-4}.
\end{equation}

Typos in some equations in \cite{Kinoshita03} 
are corrected in this article.

\section*{Acknowledgment}

This work is supported in part by the U. S. National Science Foundation under the Grant No. NSF-PHY-1316222.

\end{document}